\documentclass[11pt, a4paper]{article}
\usepackage{jheparxiv}
\usepackage[latin1]{inputenc}
\usepackage{amsmath}
\usepackage{amsfonts}
\usepackage{amssymb}
\usepackage{latexsym}
\usepackage{mathrsfs}
\usepackage{graphicx}
\usepackage{color}
\usepackage{slashed}
\usepackage{twistor}

\renewcommand{\d}{\mathrm{d}}

\subheader{\hfill \texttt{DAMTP-2014-29}}

\title{Perturbative Gravity at Null Infinity}

\author{Tim Adamo, Eduardo Casali and David Skinner}

\affiliation{Department of Applied Mathematics \& Theoretical Physics \\
        University of Cambridge \\
        Wilberforce Road \\
        Cambridge CB3 0WA, United Kingdom}

\emailAdd{[t.adamo, e.casali, d.b.skinner]@damtp.cam.ac.uk}

\abstract{We describe a theory that lives on the null conformal boundary $\scri$ of asymptotically flat space-time, and whose states encode the radiative modes of (super)gravity. We study the induced action of the BMS group, verifying that the Ward identity for certain BMS supertranslations is equivalent to Weinberg's soft graviton theorem in the bulk. The subleading behaviour of soft gravitons may also be obtained from a Ward identity for certain superrotation generators in the extended BMS algebra proposed by Barnich \& Troessaert. We show that the theory computes the complete classical gravitational S-matrix, perturbatively around the Minkowski vacuum.}

\notoc 

\begin{document}

\maketitle\vfill

\section{Introduction}

The conformal boundary of a four dimensional asymptotically flat space-time is a null hypersurface $\scri$, whose past and future components $\scri^\pm$ are topologically $\R\times S^2$~\cite{Penrose:1962ij, Penrose:1965am}.  The symmetry group of each of $\scri^\pm$ is a copy of the infinite dimensional BMS group~\cite{Bondi:1962px, Sachs:1962wk}. The BMS group is the asymptotic symmetry group of the bulk space-time and, as in AdS/CFT, we should expect it to play an important role in any candidate holographic description of gravity.  Indeed, this perspective was taken well before the advent of AdS/CFT, for instance in Ashtekar's asymptotic quantization programme~\cite{Ashtekar:1981bq, Ashtekar:1987} which encodes bulk gravitational degrees of freedom in terms of geometric data defined intrinsically on $\scri$. 

Much subsequent research has followed this general line of thought (often with the language of holography), seeking to determine the symmetry properties required for a boundary theory in asymptotically flat space-time ({\it c.f.}, \cite{Arcioni:2003td, Barnich:2009se, Bagchi:2010eg, Bagchi:2010zz, Duval:2014uva}).  Most recently, Strominger~\cite{Strominger:2013jfa} has shown that in space-times where space-like infinity is sufficiently well-behaved~\cite{Christodoulou:1991cr,Christodoulou:1993uv}, one can identify a diagonal action of the BMS groups on $\scri^+$ and $\scri^-$; this diagonal action is a symmetry of the gravitational S-matrix. In particular, the Ward identity associated with certain carefully chosen BMS generators is equivalent~\cite{He:2014laa} to Weinberg's soft graviton theorem~\cite{Weinberg:1965nx} in the bulk. It has further been suggested that the subleading behaviour of soft gravitons~\cite{White:2011yy,Cachazo:2014fwa} may be due to a Ward identity for an extension of the BMS group proposed by Barnich \& Troessaert~\cite{Barnich:2010eb}. 

This kinematic work is important because of its universality: the soft graviton theorem holds irrespective of the matter content of the theory, and receives no quantum corrections to all orders in perturbation theory\footnote{Whether or not quantum corrections to the sub-leading term arise depends subtly on whether the graviton is taken to become soft before or after expanding the regularized loop integral in $4-2\epsilon$ dimensions~\cite{Bern:2014oka,He:2014bga,Cachazo:2014dia}.}. 

\medskip

We wish to go beyond these purely kinematic considerations. The most obvious, diffeomorphism invariant observable in an asymptotically flat space-time is the S-matrix. Indeed, the S-matrix is almost tautologically holographic, being defined in terms of how states look in the distant past and future. For massless particles, the relevant asymptotic region is $\scri$, and one might hope that correlation functions in a boundary theory on $\scri$ can compute scattering amplitudes in the bulk. In fact, the usual on-shell momentum eigenstates considered in scattering amplitudes are extremely closely related to local insertions on $\scri$, as we review in section~\ref{Sec:Geom}. The precise form of the scattering amplitudes of course depends on the details of the quantum gravity in the bulk, but it seems reasonable to expect that there should exist a regime where classical (super)gravity is a good bulk description.

Traditionally, amplitudes have been computed using Feynman diagrams to evolve fields through the bulk, or else by considering a string theory whose worldsheet is mapped to a minimal surface in the bulk space-time. In recent years however, powerful techniques have been developed that compute amplitudes purely using on-shell quantities: notions such as a space-time Lagrangian or off-shell propagator do not arise. Furthermore, in these methods the building blocks from which amplitudes are constructed do not have a straightforward bulk space-time interpretation. We would like to hope that such methods provide insight into the structure of a putative holographic dual to gravity in asymptotically flat space-time, at least in some limiting regime.

\medskip

The purpose of this paper is to construct a field theory that lives entirely on (complexified) $\scri$, and whose states encode the asymptotic radiative modes of gravity in the bulk. We study the action of the BMS group on this theory, and in particular recover the Ward identity of~\cite{Strominger:2013jfa} for the action of charges associated with supertranslations, and hence the soft graviton theorem.  The theory also accommodates charges for the superrotations of the extended BMS group~\cite{Barnich:2010eb}, and when acting on correlation functions, these produce the sub-leading gravitational soft factor found in~\cite{Cachazo:2014fwa}.  Finally, we show that the simplest correlation functions of this theory produce the tree-level S-matrix of $\cN=8$ supergravity. 

Let us emphasize immediately that we do {\it not} view this theory as a realization of a boundary theory dual to gravity in asymptotically flat space.  Rather,  it may provide a perturbative description of such a theory in a regime where classical supergravity is valid in the bulk.  Nevertheless, it still provides a dynamical realization of a theory defined entirely on $\scri$ which produces bulk observables and carries a natural action of the BMS group.

The contents of the paper are as follows. In section~\ref{Sec:Geom} we provide a brief review of the geometry of $\scri$ and its complexification, as well as the asymptotic radiative modes of gravity and the BMS group.  We define our model in section~\ref{Sec:Model}, describing its BRST charge and states.  Section~\ref{Sec:Sym} discusses the realization of the (extended) BMS group in terms of charges in the model; we show how these act on correlation functions and recover the results of \cite{Strominger:2013jfa, He:2014laa, Cachazo:2014fwa}.  In section~\ref{Sec:Amp}, we evaluate the simplest correlation functions explicitly, and show that they produce the tree-level scattering amplitudes of supergravity.  We conclude in section~\ref{Sec:Concl}.

%%%%%%%%%%%%%%%%%%%%%%%%%%%%%%%%%%%%%%%%%%%%%%%%%%%%%%%%%%%
%%%%%%%%%%%%%%%%%%%%%%%%%%%%%%%%%%%%%%%%%%%%%%%%%%%%%%%%%%%

\section{The Geometry of $\scri$}
\label{Sec:Geom}

In four dimensions, the conformal boundary $\scri$ of an asymptotically flat space-time is a null hypersurface in the conformally re-scaled metric, composed of two disjoint factors $\scri=\scri^{-}\cup\scri^{+}$.  Each of $\scri^\pm$ has the topology of a light cone $\scri^{\pm}\cong \R\times S^{2}$~\cite{Penrose:1962ij, Penrose:1965am}.  Null infinity is the natural holographic screen on which the S-matrix of massless states may be defined. In Lorentzian signature, the scattering process evolves initial data on $\scri^-$ to final data on $\scri^+$. 

For the most part in this paper we shall not consider $\scri$ itself, but rather its {\it complexification} $\scri_{\C}$~\cite{PenroseBothVols}. There are many physically interesting situations in classical relativity where one \emph{must} complexify $\scri$ in order to obtain non-trivial information ({\it c.f.},~\cite{Newman:1976gc, Newman:1981fn, Adamo:2009vu}), reality conditions being imposed only subsequently.  In the context of the S-matrix, crossing symmetry implies that amplitudes extend analytically to $\scri_\C$. More generally, we expect that our choice to work on $\scri_{\C}$ without reference to a future or past boundary is closely tied to the `Christodoulou-Klainerman' property of real space-times \cite{Christodoulou:1991cr, Christodoulou:1993uv}. This in particular allows one to make an identification between the generators of $\scri^{-}$ and $\scri^{+}$, thereby selecting a single copy of the BMS group to act on all asymptotic data \cite{Strominger:2013jfa}.

Complexified null infinity $\scri_\C$ is a complex three-manifold which can be charted with coordinates $(u,\zeta,\tilde{\zeta})$, where $u$ is a complex coordinate along the null generators of $\scri_{\C}$ and $(\zeta,\tilde{\zeta})$ are complex stereographic coordinates related to the usual $(\theta, \phi)$ by $\zeta=\e^{\im\phi}\cot(\theta/2)$ and $\tilde{\zeta} = \e^{-\im\phi}\cot(\theta/2)$. (Note that $(\zeta,\tilde{\zeta})$ are not necessarily complex conjugates if $(\theta,\phi)$ are not assumed real.)  Equivalently, we can view the complexified space of generators as the product $\CP^1\times\CP^1$ of two Riemann spheres, described  by homogeneous coordinates $\lambda_{\alpha}=(\lambda_{0},\lambda_{1})$ and $\tilde{\lambda}_{\dot\alpha}=(\tilde{\lambda}_{\dot0},\tilde{\lambda}_{\dot1})$, respectively. Hence, we can chart $\scri_{\C}$ with `projective' coordinates $(u,\lambda,\tilde{\lambda})$, defined up to the equivalence~\cite{Eastwood:1982}
\begin{equation*}
 (u,\lambda,\tilde{\lambda})\sim (r\tilde{r} u, r\lambda, \tilde{r}\tilde{\lambda}), \qquad r,\,\tilde{r}\in\C^{*}.
\end{equation*}
Denoting the line bundle of complex functions on $\CP^{1}\times\CP^{1}$ which are homogeneous of degree $m$ in $\lambda$ and degree $n$ in $\tilde{\lambda}$ by $\cO(m,n)$, this means that $\scri_{\C}$ is realized as the total space of the line bundle
\be\label{cscri}
\cO(1,1)\rightarrow\CP^{1}\times\CP^{1}.
\ee  
To recover the Lorentzian real slice one simply imposes $\tilde\lambda_{\dot\alpha} = \overline{\lambda_\alpha}$ and $u= \bar u$. Thus, the real Lorentzian cones $\scri^\pm$ may each be viewed as the total space of the bundle $\cO_{\R}(1,1)\rightarrow\CP^1$, where $\cO_{\R}(1,1)$ is the restriction of $\cO(1,1)$ to real-valued functions.

\medskip

The BMS group (the asymptotic symmetry group of asymptotically flat space-times \cite{Bondi:1962px, Sachs:1962wk}) acts naturally on $\scri$, and hence on $\scri_{\C}$ by analytic continuation.  This group is the semi-direct product
\be\label{BMS1}
\mathrm{BMS}=\mathrm{ST}\ltimes\SL(2,\C).
\ee
of an infinite dimensional Abelian group ST of \emph{supertranslations} that moves one up and down a generator of the null cone, with \emph{rotations} that are the global diffeomorphisms of the space $S^2$ of generators. In terms of the coordinates $(u,\lambda,\tilde\lambda)$, the supertranslations act as
\begin{equation}
 u\rightarrow u+\alpha(\lambda,\tilde{\lambda})\,, \qquad  \lambda\to \lambda\,, \qquad\tilde{\lambda}\rightarrow\tilde{\lambda}\,, 
\end{equation}
where $\alpha$ transforms in the same way as $u$ under a rescaling of the homogeneous coordinates, and where $\tilde\lambda = \bar\lambda$ and $\alpha$ is real in Lorentzian signature. Expanding $\alpha$ in spherical harmonics, the $\ell=0,1$ terms correspond to Poincar\'e translations. These Poincar\'e translations are a symmetry of any asymptotically flat space-time, while a generic \emph{super}translation transforms one asymptotically flat solution of general relativity to another~\cite{Newman:1966ub, Ashtekar:1987}.

An asymptotically flat Lorentzian space-time carries two copies of this BMS group, acting at $\scri^\pm$ separately. As explained in~\cite{Strominger:2013jfa}, only the diagonal subgroup can act on the S-matrix. $\scri_\C$ carries an action of (one copy of) the complexified BMS group that admits independent $\SL(2,\C)$ transformations of $\lambda$ and $\tilde\lambda$ and allows $\alpha(\lambda,\tilde\lambda)$ to be complex.

It has been suggested that the BMS group can be extended by supplementing globally well-defined $\SL(2,\C)$ rotations with \emph{any} local conformal transformations of the sphere \cite{Barnich:2009se, Barnich:2010eb, Barnich:2011ct}.  This leads to an enhanced set of (singular) rotations, known as \emph{superrotations}, on the space of null generators of $\scri_\C$, which contains two copies of the Virasoro algebra at the infinitesimal level \cite{Barnich:2011mi}.

\medskip

On $\scri_{\C}$, the radiative information from the interior of the space-time is controlled by a single complex function taking values in $\cO(-3,1)$, which we denote by $\sigma^{0}(u,\lambda,\tilde{\lambda})$.\footnote{In the Newman-Penrose formalism, this is known as the `asymptotic shear' \cite{Newman:1961qr}.}  The energy flux from the interior of the space-time is encoded in the \emph{Bondi news function} \cite{Bondi:1962px},
\be\label{BNews}
N(u,\lambda,\tilde{\lambda})=\frac{\partial\sigma^{0}}{\partial u}\equiv \dot{\sigma}^{0},
\ee
taking values in $\cO(-4,0)$.  The news function has long been regarded as fundamental to studying quantum gravity on $\scri$, since it encodes the asymptotic `radiative modes' of the gravitational field \cite{Ashtekar:1981bq, Ashtekar:1987}.  Hence, a description of scattering states at $\scri_{\C}$ should encode scattering data in terms of `insertions' of news functions.

It is worth noting that the coordinate $u$ is naturally conjugate to the `frequency' of on-shell momentum eigenstates of massless particles. To define this, in place of the standard spinor helicity variables $p_{\alpha\dot\alpha} = \Lambda_\alpha\tilde\Lambda_{\dot\alpha}$, where $\Lambda_\alpha$ and $\tilde\Lambda_{\dot\alpha}$ are defined up to $(\Lambda,\tilde\Lambda)\sim (r\Lambda,r^{-1}\tilde\Lambda)$, we take a null momentum to be 
\begin{equation*}
 	p_{\alpha\dot{\alpha}}=\omega\,\lambda_{\alpha}\,\tilde{\lambda}_{\dot\alpha}
\end{equation*}
with the equivalence $(\omega,\lambda,\tilde\lambda)\sim( r^{-1}\tilde r^{-1}\omega,r\lambda, \tilde r\tilde\lambda)$. Thus, on a (complex) Minkowski background, massless momentum eigenstates appear on $\scri_\C$ as plane waves $\e^{\im \omega u}$ of frequency $\omega$, localized along the generator of $\scri_\C$ at fixed angular location $(\lambda,\tilde\lambda)\in\CP^{1}\times\CP^{1}$.

\medskip

Finally, we note that the extension of $\scri$ required to incorporate $\cN = 2p$ extended supersymmetry is straightforward: one replaces the complexified space of null generators by $\CP^{1|p}\times\CP^{1|p}$, where each factor may now be described by homogeneous coordinates $\lambda_{A}=(\lambda_{\alpha}, \eta_{a})$ and $\tilde{\lambda}_{\dot{A}}=(\tilde{\lambda}_{\dot{\alpha}}, \tilde{\eta}_{\dot{a}})$, respectively.  The $\eta_{a},\tilde{\eta}_{\dot{a}}$ are Grassmann (anti-commuting) coordinates, with $a,\dot{a}=1,\ldots,p$.  In this paper, we will take $p=4$, corresponding to a parity symmetric treatment of $\cN=8$ supergravity in which only $\cN=4$ supersymmetry is manifest. In this setting, the news function \eqref{BNews} is replaced by a {\it news supermultiplet} $\Phi$ that takes values in $\cO(0,0)$. The first component $\phi=\left.\Phi\right|_{\eta=\tilde\eta=0}$ represents a scalar field at null infinity, while the usual news tensor and its conjugate are encoded by the coefficients of $(\eta)^{4}$ and $(\tilde\eta)^4$. The multiplet terminates with a further scalar at order $(\eta\,\tilde\eta)^4$.

We abuse notation by also using $\scri_\C$ to denote the total space of $\cO(1,1)\rightarrow\CP^{1|4}\times\CP^{1|4}$, trusting the reader to distinguish between this and the usual bosonic conformal boundary by the context.

%%%%%%%%%%%%%%%%%%%%%%%%%%%%%%%%%%%%%%%%%%%%%%%%%%%%%%%%%%%
%%%%%%%%%%%%%%%%%%%%%%%%%%%%%%%%%%%%%%%%%%%%%%%%%%%%%%%%%%%

\section{The Model}
\label{Sec:Model}

Our model is a chiral CFT describing holomorphic maps $(u,\lambda,\tilde\lambda):\Sigma\rightarrow\scri_{\mathbb{C}}$ from a Riemann sphere $\Sigma$ to the supersymmetric extension of complexified null infinity, taken to be the total space of $\cO(1,1)\to \CP^{1|4}\times\CP^{1|4}$ as above.  We expect our model to serve as a description for some effective theory on $\scri_{\C}$, analogous to a worldline formalism for a field theory.  $\Sigma$ serves as the chiral complexification of the usual worldline.   

In order to implement the $\GL(1,\C)\times\GL(1,\C)$ scaling on $\scri_{\C}$ associated with \eqref{cscri} at the level of $\Sigma$, we introduce two line bundles $\cL,\tilde{\cL}\rightarrow\Sigma$ of degree $d,\tilde{d}\geq0$, respectively.  The basic fields of the model are then
\begin{equation*}
 u\in\Omega^0(\Sigma,\cL\otimes\tilde{\cL}),\qquad \lambda_A\in\Omega^0(\Sigma,\C^{2|4}\otimes\cL), \qquad \tilde{\lambda}_{\dot A}\in\Omega^0(\Sigma,\C^{2|4}\otimes\tilde{\cL}),
\end{equation*}
which describe the pullbacks to $\Sigma$ of homogeneous coordinates on $\scri_{\C}$.  We choose the  chiral action
\be
	S_1 = \frac{1}{2\pi}\int_\Sigma w\,\dbar u + \nu^{A}\dbar\lambda_{A} + \tilde{\nu}^{\dot{A}}\dbar\tilde{\lambda}_{\dot{A}}
\ee
for these fields, where $\{w,\nu^A,\tilde\nu^{\dot A}\}$ are each (1,0)-forms on the worldsheet, with gauge charges opposite those of $\{u,\lambda_A,\tilde\lambda_{\dot A}\}$, respectively. These are Lagrange multipliers that ensure the map to $\scri_\C$ is holomorphic. We also introduce fields $\psi_A$ and $\tilde\psi_{\dA}$ of opposite statistics to $\lambda_A$ and $\tilde\lambda_{\dA}$, together with their conjugates $\bar\psi^A$ and $\bar{\tilde\psi}^{\dA}$, respectively. Each of these fields is a worldsheet spinor, neutral under both $\GL(1,\C)$ scalings. Their action is
\be
	S_2 = \frac{1}{2\pi}\int_\Sigma \bar{\psi}^{A}\dbar\psi_A + \bar{\tilde\psi}^{\dot{A}}\dbar\tilde\psi_{\dot{A}}
\ee
and the combined action $S_1+S_2$ is invariant under the fermionic transformations 
\be\label{nonsusy}
	\delta\psi_{A} = \varepsilon_1\lambda_A\,,\qquad \delta\bar\psi^{\alpha} = \varepsilon_2\,\epsilon^{\alpha\beta}\lambda_\beta\,,
	\qquad
	\delta\nu^A = \varepsilon_1 \bar\psi^A - \varepsilon_2\delta^A_{\ \alpha} \epsilon^{\alpha\beta}\psi_{\dot\alpha}\,, 
\ee
with similar transformations for the tilded fields. All other fields remain invariant.

To gauge these fermionic symmetries we include bosonic ghosts $s^{1,2}\in\Omega^0(\Sigma,K^{1/2}\otimes\cL^{-1})$ and $\tilde s^{1,2}\in \Omega^0(\Sigma,K^{1/2}\otimes\tilde\cL^{-1})$ together with their antighosts $r_{1,2}$ and $\tilde{r}_{1,2}$. We also include fermionic ghosts ${\rm n},\, \tilde{\rm n}\in \Omega^0(\Sigma)$ and antighosts ${\rm m},\,\tilde{\rm m}\in \Omega^0(\Sigma,K)$ associated to gauging the $\GL(1,\C)\times \GL(1,\C)$ transformations. The ghost action is
\be
	S_3 =\frac{1}{2\pi}\int_\Sig r_a\dbar s^a + \tilde r_a \dbar \tilde s^a + \mathrm{m}\,\dbar \mathrm{n}+ \tilde{\mathrm{m}} \,\dbar \tilde{\mathrm{n}} \,.
\ee
The final ingredient is a conjugate pair of fermionic fields $\xi\in\Omega^0(\Sigma,(\cL\otimes\tilde\cL)^{-1})$ and $\chi\in\Omega^0(\Sigma,K\otimes\cL\otimes\tilde\cL)$ with action
\be
	S_4=\frac{1}{2\pi}\int_\Sigma \chi\,\dbar\xi\ .
\ee
The role of these fields will be explained below.

We take the BRST operator to be\footnote{Here we use the standard notation $\la ab\ra=\epsilon_{\alpha\beta}a^{\alpha}b^{\beta}$, $[\tilde{a}\tilde{b}]=\epsilon_{\dot{\alpha}\dot{\beta}}\tilde{a}^{\dot\alpha}\tilde{b}^{\dot\beta}$.}
\be\begin{aligned}\label{BRST}
Q_{\rm BRST}&=\oint -\mathrm{n}(w \, u+\nu^{A}\lambda_A+r_a s^a+\chi\,\xi)-\tilde{\mathrm{n}}(w\,u+\tilde{\nu}^{\dot{A}}\tilde{\lambda}_{\dot{A}}+\tilde r_a\tilde s^a+\chi\,\xi)\\
&\hspace{2cm}+s^1\lambda_{A}\bar{\psi}^A+s^2 \left\la\lambda\,\psi\right\ra+\tilde s^1\tilde{\lambda}_{\dot{A}}\tilde{\bar{\psi}}^{\dot{A}}+\tilde s^2 [\tilde{\lambda}\,\tilde\psi]\ .
\end{aligned}
\ee
This includes gaugings of the fermionic symmetries above as well as the gaugings associated to $\cL$ and $\tilde\cL$.  It is straightforward to check that $Q_{\rm BRST}$ is nilpotent and anomaly free.  For example, there is a potential $\GL(1,\C)$ anomaly $\mathfrak{a}_{\GL(1)}$ associated with the line bundle $\cL$.  This is given by the sum of squares of the fields' $\GL(1,\C)$-charges, weighted by a sign for their respective statistics:
\begin{equation*}
 \mathfrak{a}_{\GL(1)}=\sum_{i}(-1)^{F_i}\,q^2_{i}= 1_{wu}+(2-4)_{\nu\lambda}+2_{rs}-1_{\chi\xi}=0\,.
\end{equation*}
The anomalies associated with $\tilde{\cL}$ and $\cL\times\tilde{\cL}$ vanish by identical calculations.  For completeness, we also note that the central charge of this chiral CFT is given by\footnote{Non-vanishing $\mathfrak{c}$ corresponds to a non-vanishing Virasoro anomaly, but as there is no gauging of gravity on $\Sigma$ (\textit{i.e.}, no $bc$-ghost system) the role of such an anomaly is unclear.}:
\begin{equation*}
 \mathfrak{c}=2_{wu}+3(2-4)_{\nu\lambda,\,\bar{\psi}\psi}+3(2-4)_{\tilde{\nu}\tilde{\lambda},\,\tilde{\bar{\psi}}\tilde{\psi}}-4_{rs,\,\tilde{r}\tilde{s}}-4_{\mathrm{mn},\,\tilde{\mathrm{m}}\tilde{\mathrm{n}}}-2_{\chi\xi} =-20\,.
\end{equation*}

\medskip

BRST closed vertex operators are built out of the worldsheet fields in such a way as to have vanishing charge under $\cL$ and $\tilde\cL$, and be invariant under the fermionic transformations~\eqref{nonsusy}. The simplest such operators are gauge-invariant functions $\Phi=\Phi(u,\lam,\tlam)$ of the target space coordinate fields that have vanishing worldsheet conformal weight.  Crucially, these operators encode the energy flux through $\scri_{\C}$, so their correlation functions should contain information about the bulk space-time.  Expanding $\Phi$ in the fermionic coordinates on $\scri_{\C}$ gives
\be\label{vertop}
\Phi_{(0,0)}(u,\lam,\tlam)=\phi_{(0,0)}+\cdots+(\eta)^4N_{(-4,0)}+\cdots+(\teta)^4\tilde N_{(0,-4)}+\cdots+(\eta)^4(\teta)^4\tilde\phi_{(-4,-4)}\,,
\ee
where subscripts denote weights with respect to $(\lam,\tlam)$ and the component fields are functions only of the bosonic coordinates. In particular, $N_{(-4,0)}$ represents the Bondi news function \eqref{BNews}, encoding the radiative data of a negative helicity graviton, while $\tilde N_{(0,-4)}$ is the news function for the positive helicity graviton.  The other components represent analogous `news functions' for the other particle content of $\cN=8$ supergravity; for instance the 28 components with two more $\eta$s than $\tilde\eta$s represent negative helicity photons, while the 70 components with equal numbers of $\eta$ and $\tilde\eta$s are scalars\footnote{Encoding the $\cN=8$ gravitational multiplet in this way breaks the $\SU(8)$ R-symmetry group to a $\SU(4)\times\SU(4)$ subsector (in Lorentzian signature) where the two factors are related by parity symmetry.}.
We note that the vertex operators of our worldsheet CFT are not constrained to have vanishing conformal weight, so there will be infinite towers of states beyond these simplest ones. We will see below that correlation functions of arbitrarily many $\Phi(u,\lam,\tlam)$ operators do not excite these other states. It would be interesting to explore their meaning in detail.

\medskip

The bosonic antighost fields $r,\tilde r$ have zero modes when $d,\td>0$.  To make sense of the path-integral measure we must insert picture changing operators (PCOs) to absorb these zero modes:
\be\label{PCO}
 \Upsilon = \delta(r_1)\delta(r_2)\,\lam_A\psib^A \left\la\lam\,\psi\right\ra, \qquad 
 \tilde\Upsilon=\delta(\tilde r_1)\delta(\tilde r_2)\,\tlam_{\dA}\bar{\tilde\psi}^{\dA}\; [\tlam\,\tilde\psi].
\ee
Insertions of $\Upsilon$ ($\tilde\Upsilon$) absorb $2d$ ($2\tilde d$) zero modes of the $r_1,r_2$ ($\tilde r_1,\tilde r_2$) antighosts. As usual, the correlation function doesn't depend on the location of the PCOs.

We must now pick a measure with which to integrate over the moduli space of vertex operator locations. This will reveal the role of the $\xi\chi$ system. Consider the composite operator $w\,\chi$. This is an uncharged fermionic quadratic differential on the worldsheet, and is BRST closed\footnote{In particular $\{Q_{\rm BRST}, w\,\chi\} \neq T$, so $w\,\chi$ cannot be interpreted as a composite $b$ ghost.}. In the presence of vertex operators at points $\{x_1,\cdots,x_n\}\in\Sigma$, it has $n-3$ zero modes.  As usual in string theory, if $\{\mu_j\}$ form a basis of Beltrami differentials on the punctured worldsheet, we can construct a top holomorphic form on the moduli space of these punctures as $\prod_{j=1}^{n-3}\,(w\,\chi|\mu_j)$, the bracket denoting integration over $\Sigma$. This choice of measure places an important constraint on the possible degrees of the line bundles $\cL$ and $\tilde\cL$. Since  $\chi$ has $d+\tilde d-1$ zero modes, the correlation function vanishes unless
\be\label{constraint}
	d+\tilde{d}=n-2,
\ee 
As in~\cite{Witten:2004cp}, this amounts to the requirement that $\cL\otimes\tilde\cL \cong K(x_1+\cdots+x_n)$.

\medskip

The simplest correlation function in our model is thus
\be
\label{correlator}
 \cM_{n,d}=\left<\prod_{i=1}^n\Phi(\sig_i)\prod_{j=4}^n\left(w\,\chi|\mu_j\right)\prod_{k=1}^d\Upsilon_k\;\prod_{l=1}^{\tilde d}\tilde\Upsilon_l\right> 
 	=\left<\!\!\!\!\left<\prod_{i=1}^3\Phi(\sig_i)\prod_{j=4}^n\int_{\Sigma}\chi(\sig_j)\dot{\Phi}(\sigma_j)\right>\!\!\!\!\right>,
\ee
where $\dot\Phi=\partial_{u}\Phi$ takes values in $\cO(-1,-1)$ on $\scri_{\C}$ and $\langle\!\langle\cdots\rangle\!\rangle$ denotes a correlator in the presence of the PCOs.

%%%%%%%%%%%%%%%%%%%%%%%%%%%%%%%%%%%%%%%%%%%%%%%%%%%%%%%%%%
%%%%%%%%%%%%%%%%%%%%%%%%%%%%%%%%%%%%%%%%%%%%%%%%%%%%%%%%%%

\section{Symmetries and Ward Identities}
\label{Sec:Sym}

We expect that all the symmetries of $\scri_{\C}$ have a realization in terms of charges on $\Sigma$, which act on correlators such as \eqref{correlator}.  The simplest such symmetry is the Poincar\'e group, which is generated by the charges
\be\label{Poincare}
Q_{\SL(2,\C)}=\oint a^{\alpha}_{\ \beta}\,\lam_\al(\sig)\nu^\beta(\sig)+\mathrm{c.c.} \qquad\hbox{and}\qquad Q_{\mathrm{T}}=\oint b^{\al\dal}\,\lam_\al(\sig)\tlam_{\dal}(\sig)\,w(\sig)\,,
\ee
where $a^{\alpha}_{\ \beta}$ and $b^{\alpha\dot{\alpha}}$ are constant and the former is traceless. It is easy to see that these charges commute with the action and are \emph{bona fide} symmetries of the model, just as the Poincar\'e group is an asymptotic symmetry of every asymptotically flat space-time.

However, as discussed in section \ref{Sec:Geom}, there is a larger symmetry group on $\scri_{\C}$: the infinite dimensional BMS group, built from Lorentz transformations and supertranslations.  The latter are generated on $\Sigma$ by charges
\be\label{ST}
 Q_{\mathrm{ST}}=\oint f(\lam,\tlam)\,w(\sig),
\ee
where $f$ is a function of weight one in both $\lam$ and $\tlam$.\footnote{$Q_{\mathrm{ST}}$ generates `complexified' supertranslations on $\scri_{\C}$; to restrict to the real $\scri$, we set $\tilde{\lambda}=\bar{\lambda}$ so that $f$ becomes a smooth function on $\CP^{1}$, defining a real supertranslation.} Unlike the Poincar\'e charges \eqref{Poincare}, the general supertranslation charge \eqref{ST} will have poles, so its commutator with the action will be non-vanishing at these poles.  This is expected, since our realization of $\scri_{\C}$ as a vector bundle over $\CP^{1}\times\CP^{1}$ endows it with more structure than necessary.  In particular, the choice of an origin for this vector bundle is equivalent to a choice of classical vacuum from the perspective of asymptotic quantization \cite{Ashtekar:1981bq, Ashtekar:1987}; since supertranslations map one vacuum to an inequivalent vacuum, we shouldn't expect them to be exact symmetries of our model.  Nonetheless, $Q_{\rm ST}$ \emph{does} give nontrivial information in the form of a Ward identity containing information about soft gravitons \cite{Strominger:2013jfa, He:2014laa}.  

In particular, consider the supertranslation given by the charge
\be\label{qsoft}
Q^{(1)}_{\mathrm{ST}}=\oint f^{(1)}(\lambda,\tilde{\lambda})\,w(\sigma), \qquad f^{(1)}(\lambda,\tilde{\lambda})= \frac{a^\al b^\beta\tlam_{s}^{\dal}}{\left<a\,s\right>\left<b\,s\right>}\frac{\lam_\al(\sig)\lam_\beta(\sig)\tlam_{\dal}(\sig)}{\left<s\,\lam(\sig)\right>},
\ee
where $(\lambda_{s},\tilde{\lambda}_{s})$ is a fixed point on the space of generators of $\scri_{\C}$ associated with the insertion of a soft graviton. Note that $f^{(1)}$ has weight (1,1) in $(\lambda(\sigma),\tilde\lambda(\sigma))$ as required for a supertranslation, and weight $(-3,1)$ in $(\lambda_s,\tilde\lambda_s)$ as for the asymptotic shear of a soft graviton. Inserting this charge into \eqref{correlator}, its effect is to differentiate each vertex operator $\Phi_i$ in the $u$-direction.  Assuming that these operators are momentum eigenstates of frequency (or energy) $\omega_i$, this results in a Ward identity
\be\label{softWI}
\left<\!\!\!\!\left< Q^{(1)}_{\mathrm{ST}}\;\prod_{i=1}^3\Phi(\sig_i)\prod_{j=4}^n\int_{\Sigma}\chi(\sig_j)\dot{\Phi}(\sigma_j)\right>\!\!\!\!\right> = \sum_{i=1}^{n} \omega_{i} \,f^{(1)}(\lambda(\sigma_i),\tilde{\lambda}(\sigma_i))\;\cM_{n,d}\ ,
\ee
where on the left hand side, the contour in $Q^{(1)}_{\rm ST}$ is taken along $|\langle\lambda(\sigma)\,s\rangle|=\epsilon$. This is equivalent to the Ward identity found in \cite{Strominger:2013jfa} for the supertranslations generated by \eqref{qsoft}.

More specifically, suppose that we represent the one-particle state by the explicit momentum eigenstate
\be\label{momeig}
\Phi_i=\int \frac{\rd t_i \,\rd \tilde t_i}{t_i\,\tilde t _i\, \omega_i}\, \delta^{2|4}(\lam_i-t_i\lam(\sig_i))\;\delta^{2|4}(\tlam_i-\tilde t_i\tlam(\sig_i))\; \e^{t_i\tilde t_i \omega_i\,u(\sig_i)}.
\ee
Then on the support of the delta functions in these vertex operators, the action of $Q^{(1)}_{\mathrm{ST}}$ on the correlator reads
\begin{multline}\label{softfactor}
 \sum_{i=1}^{n}\frac{a^{\alpha}b^{\beta}\tilde{\lambda}^{\dot{\alpha}}_{s}}{\la a\,s\ra\,\la b\,s\ra} t_{i}\tilde{t}_{i}\omega_{i}
 \left<\!\!\!\!\left<\frac{\lambda_{\alpha}(\sigma_i)\lambda_{\beta}(\sigma_i)\tilde{\lambda}_{\dot{\alpha}}(\sigma_i)}{\la s\,\lambda(\sigma_i)\ra} \prod_{k=1}^3\Phi(\sig_k)\prod_{j=4}^n\int_{\Sigma}\chi(\sig_j)\dot{\Phi}(\sigma_j)\right>\!\!\!\!\right> \\
=\ \sum_{i=1}^{n}\omega_{i}\frac{[s\,i]}{\la s\,i\ra}\frac{\la a\,i\ra\;\la b\,i\ra}{\la a\,s\ra\;\la b\,s\ra}\;\cM_{n,d}.
\end{multline}
This is precisely Weinberg's soft graviton theorem as re-derived in the context of supertranslations acting on the S-matrix in \cite{He:2014laa}.  In our model, the universal soft graviton factor arises from the action of a charge generating a supertranslation, which effectively creates the soft graviton at the position $(\lambda_{s},\tilde{\lambda}_s)\in\CP^{1}\times\CP^{1}$.

More general supertranslations (having additional or higher-order poles) are related to the creation of multiple soft gravitons.  Our model also includes supersymmetric extensions of supertranslations, which correspond to other soft particles in the spectrum of $\cN=8$ supergravity.  Hence, the supertranslations \eqref{ST} combined with $Q_{\SL(2,\C)}$ generate the action of the BMS group in our model.

\medskip

Interestingly, we can easily incorporate the superrotations of the extended BMS group as well.  The relevant charge on $\Sigma$ is
\be\label{superr}
Q_{\mathrm{SR}}=\oint R(\lam,\tlam)^{\al}_{\ \beta}\, \lam_\al(\sig)\,\nu^\beta(\sig)+ \tilde R(\lam,\tlam)^{\dal}_{\ \dot\beta}\, \tlam_{\dal}(\sig)\,\tnu^{\dot\beta}(\sig) \,,
\ee
where $R(\lam,\tlam)^\al_{\ \beta}$, $R(\lam,\tlam)^{\dal}_{\ \dot\beta}$ are traceless, weightless holomorphic functions of $(\lam,\tlam)$.  General operators of this form suffer from normal ordering ambiguities, but a large interesting class are free from such problems.  For instance, consider \eqref{superr} with
\begin{equation*}
 	R^{\alpha}_{\ \beta}=0, \qquad 
	\tilde R^{\dal}_{\ \dot\beta}=\frac{\langle a\,\lam(\sig)\rangle}{\langle s\, \lambda (\sigma)\rangle} \frac{\tlam_s^{\dal}\,\tlam_{s\dot\beta}}{\left<a\, s\right>}.
\end{equation*}
A calculation similar to that which led to \eqref{softfactor} gives the action of this charge on the correlator with momentum eigenstates:
\begin{equation}\label{subleading}
\left<\!\!\!\!\left< Q_{\mathrm{SR}}\;\prod_{i=1}^3\Phi(\sig_i)\prod_{j=4}^n\int_{\Sigma}\chi(\sig_j)\dot{\Phi}(\sigma_j)\right>\!\!\!\!\right> 
=  -\sum_{i=1}^{n} \frac{[s\,i]}{\left<s\,i\right>}\frac{\left< a\, i\right>}{\left<a\, s\right>}\tlam_{s\dal}\frac{\p}{\p\tlam_{i\dal}}\;\cM_{n,d} \,,
\end{equation}
where the contour for $Q_{\mathrm{SR}}$ is as before.  The last line is the holomorphic subleading soft graviton contribution recently discussed by Cachazo and Strominger \cite{Cachazo:2014fwa}.  It was conjectured that this subleading contribution is related to the action of superrotations; \eqref{subleading} demonstrates this explicitly at the level of charges acting on the correlator.

%%%%%%%%%%%%%%%%%%%%%%%%%%%%%%%%%%%%%%%%%%%%%%%%%%%%%%%%%%%
%%%%%%%%%%%%%%%%%%%%%%%%%%%%%%%%%%%%%%%%%%%%%%%%%%%%%%%%%%%

\section{Scattering Amplitudes}
\label{Sec:Amp}

Having shown that our model has vertex operators naturally encoding the asymptotic radiative degrees of freedom, and that the charges corresponding to the BMS group have a natural action on its correlators, we now turn to the evaluation of the correlation functions \eqref{correlator} themselves.   First, notice that the PCO insertions only have non-trivial Wick contractions with other PCOs of the same type.  The resulting correlation function on the PCOs can then be computed using the arguments employed for PCOs in \cite{Skinner:2013xp}, resulting in
\begin{equation}
\label{resultants}
 \left\la \prod_{k=1}^d\Upsilon_k\;\prod_{l=1}^{\tilde d}\tilde\Upsilon_l\right\ra=\mathrm{R}(\lambda)\;\mathrm{R}(\tilde{\lambda}),
\end{equation}
where $\mathrm{R}(\lambda)$, $\mathrm{R}(\tilde{\lambda})$ are the \emph{resultants} of the maps
\be\label{maps}
 \lambda_{\alpha}:\Sigma\rightarrow\CP^{1}, \qquad \tilde{\lambda}_{\dot{\alpha}}:\Sigma\rightarrow\CP^{1},
\ee
respectively \cite{Cachazo:2013zc}.  Recall that the resultant R($\lambda$) vanishes iff both $\lambda_\alpha(\sigma_*)$ vanish simultaneously for some $\sigma_*\in\Sigma$. The factor~\eqref{resultants} thus ensures that the amplitude receives contributions only when $(\lambda_\alpha(\sigma),\tilde\lambda_{\dot\alpha}(\sigma))$ is a well-defined map to $\CP^1\times\CP^1$.

Evaluating the remainder of the correlator with the choice of momentum eigenstates \eqref{momeig} for $\Phi$ leads to
\begin{multline}\label{amp1}
 \cM_{n,d}=\int \frac{\prod_{r=1}^{d+1}\d^{2|4}\lambda_{r}^{0}\,\prod_{s=1}^{\tilde{d}+1}\d^{2|4}\tilde{\lambda}_{s}^{0}}{\mathrm{vol}\left(\C^{*}\times\C^{*}\right)}\: \frac{\mathrm{R}(\lambda)\,\mathrm{R}(\tilde{\lambda})\; |\sigma_{4}\cdots\sigma_{n}|}{\prod_{j=1,2,3}t_{j}\,\tilde{t}_{j}\,\omega_j\, \D\sigma_{j}}  \\
\prod_{a=1}^{d+\tilde{d}+1}\delta\!\left(\sum_{i=1}^{n}t_{i}\tilde{t}_{i}\omega_{i}\,\mathfrak{s}_{a}(\sigma_i)\right) \prod_{i=1}^{n}\D\sigma_{i}\, \d t_{i}\, \d\tilde{t}_{i}\; \delta^{2|4}(\lam_i-t_i\lam(\sig_i))\;\delta^{2|4}(\tlam_i-\tilde t_i\tlam(\sig_i))\,.
\end{multline}
Here, the measure in the first line is over the zero modes of the maps \eqref{maps}, while the quotient by two $\C^{*}$-freedoms reflects the rescaling symmetry associated with $\cL$ and $\tilde{\cL}$. The $\sigma_{i}^{\underline{\alpha}}=(\sigma_{i}^{\underline{0}},\sigma_{i}^{\underline{1}})$ are homogeneous coordinates on $\Sigma$, which have $\SL(2,\C)$-invariant contraction $\epsilon_{\underline{\alpha}\underline{\beta}}\sigma_{i}^{\underline{\alpha}}\sigma_{j}^{\underline{\beta}}=(i\,j)$.  The Vandermonde determinant
\begin{equation*}
 |\sigma_{4}\cdots\sigma_{n}|:=\prod_{4\leq i<j\leq n}(i\,j) 
\end{equation*}
is produced by the $n-3$ $\chi$-insertions, and $\D\sigma_{i}:=(\sigma_{i}\,\d\sigma_{i})$ is the natural weight $+2$ holomorphic measure on $\Sigma$.  Finally, the first set of $\delta$-functions in the second line arises by performing the integral over zero modes for the map component $u:\Sigma\rightarrow\C$,
\begin{equation*}
 \int \d^{d+\tilde{d}+1}u^{0}\ \exp\left[\im\sum_{i=1}^{n}t_{i}\tilde{t}_{i}\omega_{i}\,u(\sigma_i)\right]
 =\prod_{a=1}^{d+\tilde{d}+1}\delta\!\left(\sum_{i=1}^{n}t_{i}\tilde{t}_{i}\omega_{i}\,\mathfrak{s}_{a}(\sigma_i)\right),
\end{equation*}
for $\{\mathfrak{s}_{a}\}$ a basis of $H^{0}(\Sigma, \cL\otimes\tilde{\cL})$.  

The expression \eqref{amp1} for $\cM_{n,d}$ can be manipulated into a more recognizable form.  Using the constraint \eqref{constraint}, one can show that (\textit{c.f.}, \cite{Roiban:2004yf, Witten:2004cp})
\begin{equation*}
 \prod_{a=1}^{d+\tilde{d}+1}\delta\!\left(\sum_{i=1}^{n}t_{i}\tilde{t}_{i}\omega_{i}\,\mathfrak{s}_{a}(\sigma_i)\right)
 =\frac{1}{|\sigma_{1}\cdots\sigma_{n}|}\int_{\C}\d r\;\prod_{i=1}^{n}\delta\!\left(t_{i}\tilde{t}_{i}\omega_{i}-\frac{r}{\prod_{j\neq i}(\sigma_i\sigma_j)}\right)\,.
\end{equation*}
Inserting this identity into \eqref{amp1} and working on the support of the various delta-functions produces an equivalent expression for the correlator:
\begin{multline}\label{amp2}
 \cM_{n,d}=\int \frac{\prod_{r=1}^{d+1}\d^{2|4}\lambda_{r}^{0}\,\prod_{s=1}^{\tilde{d}+1}\d^{2|4}\tilde{\lambda}_{s}^{0}}{\mathrm{vol}\left(\SL(2,\C)\times\C^{*}\times\C^{*}\right)}\;\mathrm{R}(\lambda)\,\mathrm{R}(\tilde{\lambda})\,\frac{\d r}{r^3}\, \prod_{i=1}^{n} \delta\!\left(t_{i}\tilde{t}_{i}\omega_{i}-\frac{r}{\prod_{j\neq i}(\sigma_i\sigma_j)}\right) \\
\times\;\D\sigma_{i}\, \d t_{i}\, \d\tilde{t}_{i}\; \delta^{2|4}(\lam_i-t_i\lam(\sig_i))\;\delta^{2|4}(\tlam_i-\tilde t_i\tlam(\sig_i))\,.
\end{multline}
Using one of the $\C^{*}$-freedoms to fix the $r$-integral, this expression is equal to a representation derived in \cite{Cachazo:2013zc} of the Cachazo-Skinner formula for the tree-level S-matrix of $\cN=8$ supergravity \cite{Cachazo:2012kg}.  Hence, we confirm that the simplest correlation function of the model, with vertex operators represented by momentum eigenstates, produces the tree-level scattering amplitudes of gravity.

%%%%%%%%%%%%%%%%%%%%%%%%%%%%%%%%%%%%%%%%%%%%%%%%%%%%%%%%%%%
%%%%%%%%%%%%%%%%%%%%%%%%%%%%%%%%%%%%%%%%%%%%%%%%%%%%%%%%%%%

\section{Conclusions}
\label{Sec:Concl} 

In this paper we have constructed a QFT living on the complexification of $\scri$. We have shown that its states include modes corresponding to the radiative data of bulk supergravity and have investigated the Ward identities corresponding to the action of the (extended) BMS group. Finally, we showed that correlation functions in this theory reproduce the classical perturbative S-matrix of $\cN=8$ supergravity in a form originally obtained in~\cite{Cachazo:2012kg,Cachazo:2013zc}. This theory can be viewed as a holographic description of gravity in asymptotically flat space-time, valid in the regime where classical supergravity is a good description in the bulk.

The boundary theory has been presented here in terms of a worldsheet description, and is thus restricted to perturbation theory. It would clearly be interesting to obtain the target space description of this theory. The model here is a close cousin of the twistor and ambitwistor models of~\cite{Skinner:2013xp,Geyer:2014fka}, which are best understood as living in the {\it asymptotic} twistor spaces of an asymptotically flat space-time. It also appears to be closely related to Schild's null strings~\cite{Schild:1976vq}, which have arisen recently as a description of massless states in theories invariant under the conformal Carroll group~\cite{Bagchi:2013bga,Duval:2014lpa}. 

An important outstanding issue is the relevance of these ideas in higher dimensions. While Weinberg's soft graviton theorem holds in arbitrary dimensions, as does the sub-leading soft behavior at least at tree-level~\cite{Schwab:2014xua,Afkhami-Jeddi:2014fia}, the asymptotic symmetry group of an asymptotically flat space-time of dimension $d>4$ does not admit supertranslations~\cite{Hollands:2003ie,Tanabe:2009va}. It is intriguing that the current, tree-level, proofs of these higher-dimensional subleading soft theorems make use of the scattering equations~\cite{Cachazo:2013hca} which arise most naturally in higher dimensional ambitwistor theory~\cite{Mason:2013sva,Adamo:2013tsa,Berkovits:2013xba}. These and other issues will be explored elsewhere.

\bigskip

\bigskip

\noindent  {\bf Acknowledgements:} E.C. and D.S. would like to thank F. Cachazo for helpful discussions. The work of T.A. is supported by a Title A Research Fellowship at St. John's College, Cambridge. The work of E.C. is supported in part by the Cambridge Commonwealth, European and International Trust. The work of D.S. in supported in part by a Marie Curie Career Integration Grant. The research leading to these results has received funding from the European Research Council under the European Community's Seventh Framework Programme (FP7/2007-2013) / ERC grant agreement no. [247252] and Marie Curie grant agreement no. [631289].

\bibliography{scri}
\bibliographystyle{JHEP}

\end{document}